\documentclass[prb,preprint]{revtex4-1} 

\usepackage{graphicx}
\usepackage{dcolumn}
\usepackage{bm}
\usepackage{setspace}

\begin{document}


\title{Study of Spherical Array Target for Long-Baseline Neutrino Experiment}

\author{Quynh M. Nguyen}

\affiliation{University of Minnesota, Minneapolis, MN, USA, 55455}
\affiliation{Fermi National Accelerator Laboratory (FNAL), Batavia, IL, USA,60510}
\author{Alberto Marchionni}
\affiliation{Fermi National Accelerator Laboratory (FNAL), Batavia, IL, USA,60510}

\date{August 8, 2014}

\begin{abstract}
Spherical Array Target was studied by implementing the geometry in LBNE's Beam Simulation source code g4lbne version v3r2p4 and Monte Carlo. To compare with Nominal LBNE target, unoscillated Far Detector neutrino flux was produced using different parameters: sphere diameter 17mm and 13mm, different longitudinal positions, two interaction length and less, beam size R/3 and 1.7mm, and beam offset from 50 $\mu$m to 1mm. The 1.86 interaction length (901mm), 17mm diameter target, with beam size 1.7mm gives higher $\nu_{\mu}$ flux up to 10\% from 0-3.5 GeV, and suppresses flux up to 70\% at energy higher than 3.5 GeV.

\end{abstract}

\maketitle

%

\section{\label{sec:level1}Introduction to LBNE and the Target}
The Long-Baseline Neutrino Experiment (LBNE) collaboration plans a comprehensive experiment that will study CP violation, neutrino mass hierarchy and other physics beyond the Standard Model\cite{lbne} using a high intensity 1300km baseline accelerator neutrino beam and an advanced 34kton liquid argon TPC as the far detector. 

The LBNE Beamline at Fermilab will be designed to provide a neutrino beam of sufficient intensity and proper energy range to meet the goals of the experiment. The design is a conventional, horn-focused beamline \cite{volume2}. Proton beam from Fermilab Main Injector will be extracted and transport to a target area where collisions generate a beam of charged particles mainly pions and kaons. The charge particles will be focused by magnetic horns, lead into decay-pipe tunnel where they decay to generate neutrino beam, aiming toward the Far Detector. 

The current nominal target is a segmented graphite target, such as one used by NuMi. The spherical array target studied and designed by Rutherford Appleton Laboratory (RAL) promises interesting mechanical properties but there was no study on its ability to produce neutrino flux at the Far Detector. This paper presents a study of the spherical array target (SAT) for the LBNE's beam-simulation group. Geometry of the new target was developed in LBNE Geant4's derivative, g4lbne version v3r2p4.


\section{Design of SAT and Study Motivation}

\begin{figure}
\includegraphics[scale=0.6]{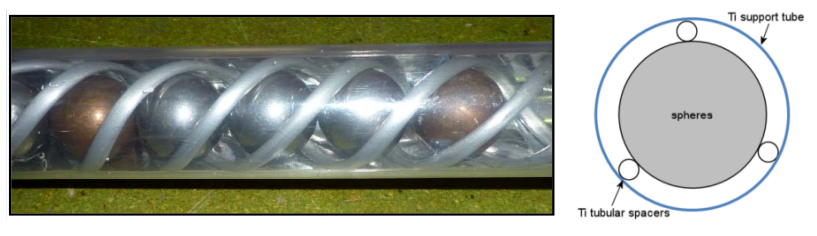} 
\caption{\cite{ral}Concept design of an array of spheres with triple helix spacers and Perspex outter casing}
\label{sat}
\end{figure}
As designed by RAL, the target is an array of spheres, containing in a helical spacer and a Titanium support cylindrical tube (Fig. \ref{sat}). The spheres are made of Beryllium. Properties of graphite and beryllium are provide in Table. \ref{properties}. Except a higher modulus of elasticity, Beryllium generally offers better mechanical properties as a target material than graphite such as higher specific heat, thermal conductivity and yield strength.

\begin{figure}
\includegraphics[scale=0.4]{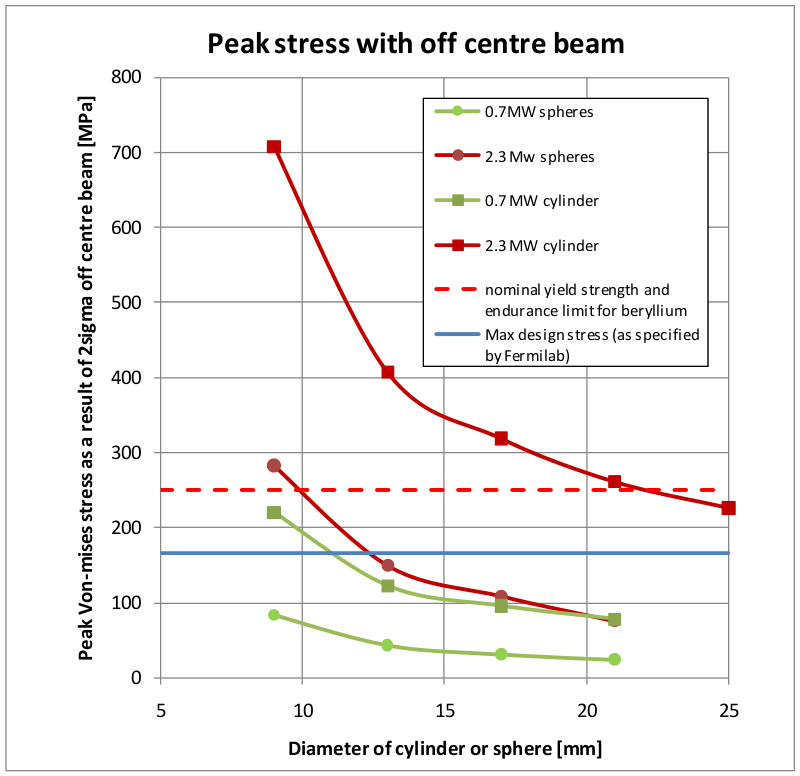} 
\caption{\cite{ral}Peak stress for a 2 sigma off centre beam. Study done by RAL}
\label{peakStress}
\end{figure}

The spherical design is to avoid sharp edges and to offer most homogeneous stress field. Furthermore, spheres also have no structural weak point \cite{ral} and provide more surface for cooling. Stress simulations by RAL have shown that, an array of spheres of diameter as small as 13mm can keep the mechanical stress below the design point at beam power 2.3MW. (Fig. \ref{peakStress}). 
\begin{figure}
\includegraphics[scale=0.7]{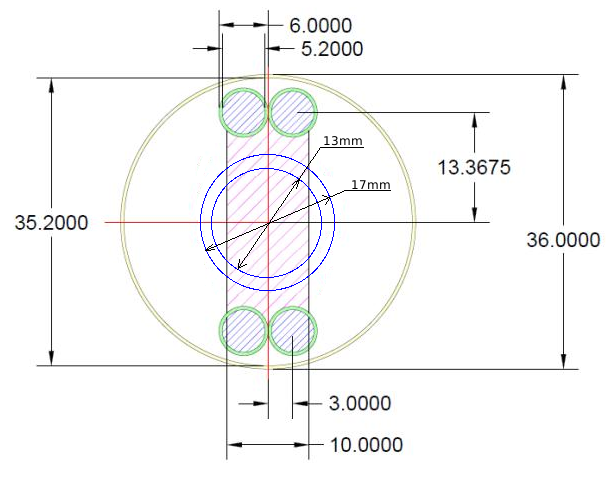} 
\caption{Dimension of sphere targets (13mm and 17 mm) compared to NuMi target. Helium can size used in this study is 36mm.}
\label{numi}
\end{figure}
Thus, spheres diameter of 13mm and 17mm were chosen. The diameter of a simple Helium can was inferred from T2K's target \cite{t2k} to be approximately 36mm. The can is made of Titanium 0.4mm thick.

\begin{table}[h]
\centering
\begin{tabular}{l*{1}{c}r}
\hline
\hline
            &           Graphite (POCO ZXF-5Q)  & Beryllium (S-65C)  \\
\hline
    Apparent density (g cm$^{-3}$)    & 1.81      & 1.848\footnote{Particle Data Group}     \\        
    Compressive Strength (MPa)      & 195       & 260 \\
    Tensile Strength (MPa)          & 90        & 370 \\
    Yield Strength (MPa)            &  -         &  240 \footnote{GoodFellow.com}   \\
    Modulus of Elasticity (GPa)     & 12.5      & 310 \\
    Thermal Conductivity (W m$^{-1}$ K$^{-1}$)   &70     &200 \\
    Coeff. of Thermal Expansion ($\mu$m m$^{-1}$ K$^{-1}$) & 8.1 & 10.7\\
    Specific Heat (J kg$^{-1}$ K$^{-1}$)      & 710  & 1770\\
    Nuclear interaction length $\lambda_I$ \footnote{Particle Data Group}(g cm$^{-2}$) & 85.8   & 77.8  \\

\hline
\hline
\end{tabular} 
\caption{\cite{volume2} Properties of graphite and beryllium at 20$^{\circ}$C, from the manufacturers, except as noted}
\label{properties}
\end{table}

\section{\label{sec:level1}Implementation}
\subsection{Target length calculation}
\begin{figure}
\frame{\includegraphics[scale=0.3]{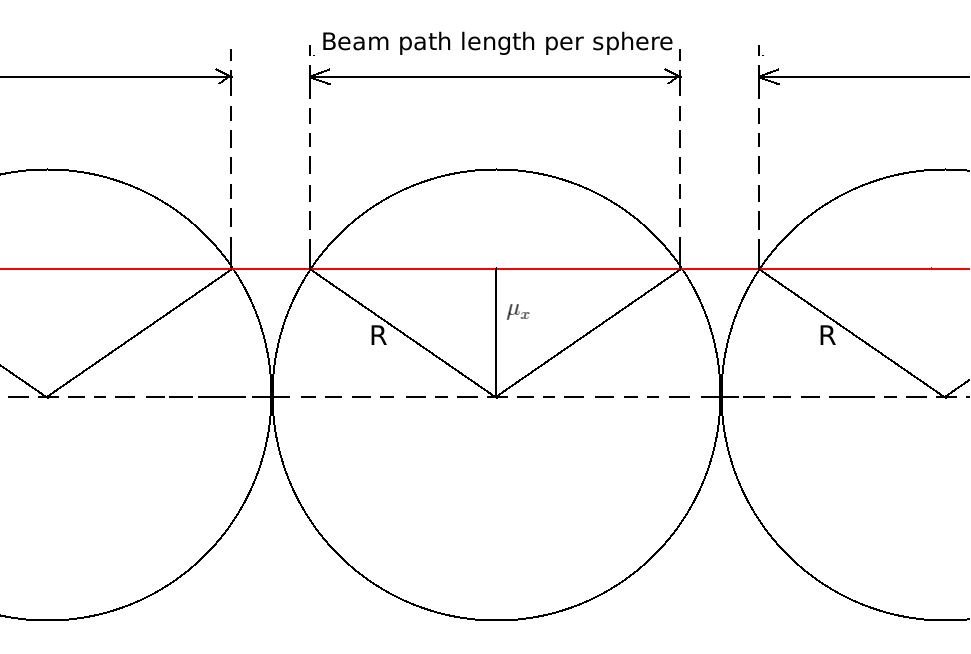}}
\caption{Path length of off-centre beam. Off-centre beam passes through less material. Thus the target is longer than a simple cylinder to make two effective interaction lengths.}
\label{path}
\end{figure}

The length of SAT was calculated such that on average, the beam traversing SAT longitudinally has a path length of two interaction lengths. The path length of proton beams depends on beam's offset from target's center, and the beam is assumed to have a bi-dimensional Gaussian distribution center at target longitudinal axis with sigma equal 1/3 of the sphere radius, which means 99.89\% of the beam is within the cross-section (However, smaller beam size was also studied). The effective interaction length can be expressed as:
\begin{equation}
N \times \int_{-\sqrt{R^2-x^2}}^{\sqrt{R^2-x^2}} \int_{-R}^R \! \left(\frac{1}{{2\pi}\sigma_x\sigma_y}e^{-\frac{x^2}{2\sigma_x^2}-\frac{y^2}{2\sigma_y^2} }\right)\times2\sqrt{R^2-x^2-y^2}\, \mathrm{d}x \mathrm{d}y = 2 \times \text{interaction length}
\end{equation}
where $2\sqrt{R^2-x^2-y^2}$ is the parallel path length of a beam traversing through a sphere (see Fig. \ref{path}), N is the number of spheres in the target. Here, both $\sigma_x$ and $\sigma_y$ equal to R/3. Beryllium interaction length is 421mm, calculated by the ratio Nuclear Interaction length/Density from Table. \ref{properties}. Solving for number of spheres give 57 spheres (rounded) of diameter 17mm (total length 969mm) and 74 spheres of diameter 13mm (total length 975mm). In the source code, a tolerance of 5 microns is also added as allowed gap between each spheres.

\subsection{Probability of interaction of off-axis proton}
When the beam is offset, it traverses less target material, so does the probability of interaction. For a Gaussian beam distribution, the probability of interaction for off-centre beam is:
\begin{equation}
P (\mu_x,\mu_y) = \int_{-\sqrt{R^2-x^2}}^{\sqrt{R^2-x^2}} \int_{-R}^R \! \left(\frac{1}{{2\pi}\sigma_x\sigma_y}e^{-\frac{(x-\mu_x)^2}{2\sigma_x^2}-\frac{(y-\mu_y)^2}{2\sigma_y^2} }\right)\times \left( 1- e^{-\frac{2N\sqrt{R^2-x^2-y^2}}{\text{interaction length}}}\right) \, \mathrm{dx} \mathrm{dy}
\label{prob_inter}
\end{equation}
In study of beam position using Monte Carlo simulations, beam offset was set in x direction (y direction would give the same result regarding the simulation). To check on the Monte Carlo, the integral in Eq. \ref{prob_inter} with $\mu_x$ as beam offset, $\mu_y = 0$, is used to approximate $P (\mu_x,\mu_y)$, in comparison to on-centre beam. Relative magnitude of P $(\mu_x,\mu_y)$ should agree with neutrino flux.
\begin{table}[h]
\centering
\setlength{\tabcolsep}{3.75pt}
\begin{tabular}{l*{7}{c}r}
\hline
\hline
 Beam offset (mm)  & 0      &   0.05  & 0.1 & 0.2  &   0.4 &  1  & 2 & 3\\
  \hline
    
Probability of & 85.043\% & 85.042\% & 85.04\% & 85.02\% & 84.95\% & 84.47\% & 82.63\% & 79.27\% \\
interaction (\%) & &  &  &  &  &  &  \\
    \hline
Ratio to  & 1 & 0.99998 & 0.99993 & 0.99973 & 0.99893 & 0.99320 & 0.97163 & 0.93215  \\  
on-centre beam &  & &  &  & &  &   \\  
\hline
\hline
\end{tabular} 
\caption{Probability of interaction of protons hitting the target at different offset from the target center}
\label{prob}
\end{table}
\subsection{Code development}
\begin{figure}
\includegraphics[scale=0.38]{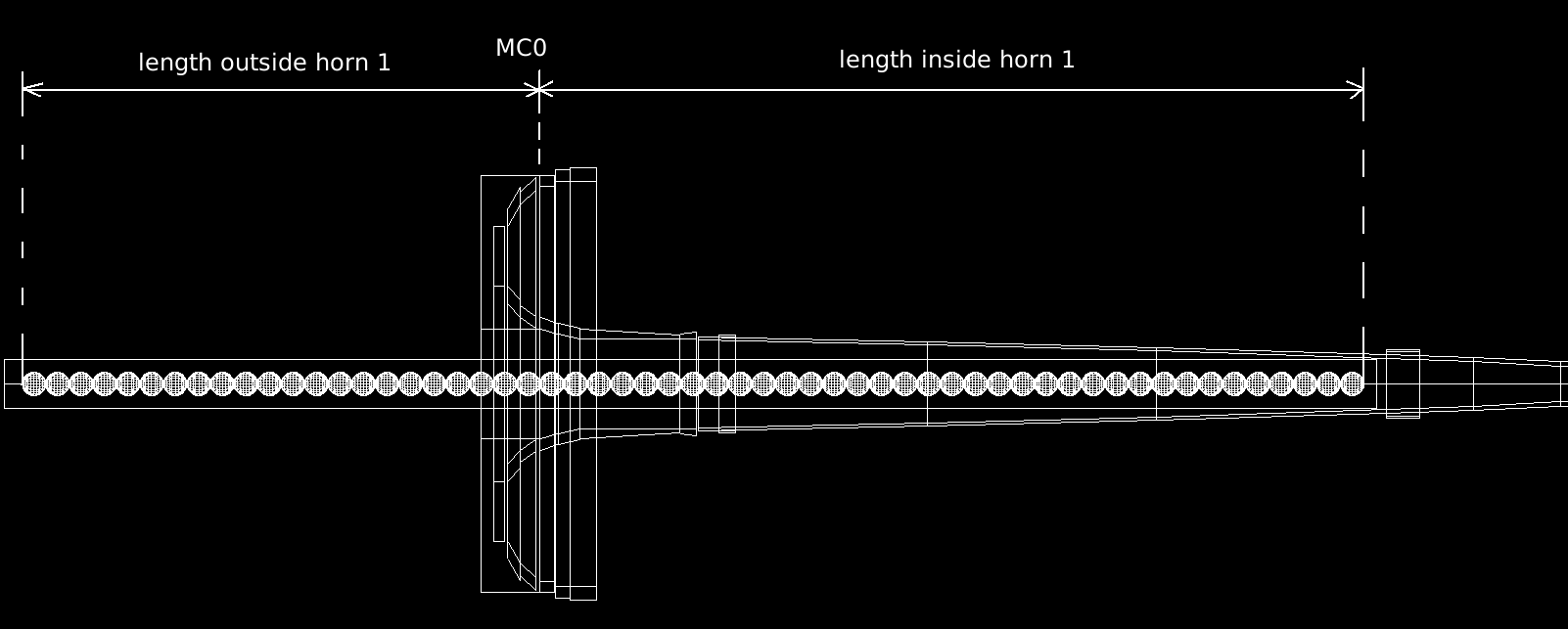} 
\caption{HepRApp visualizaion of horn 1 (inner conductor) and spherical array target }
\label{horn1}
\end{figure}

For coding convenience, the target is divided into two volumes, upstream and downstream, with the break at MC0 of horn 1 (see Fig. \ref{horn1}). In this initial study, the code only allows target position to change in integer steps of sphere diameter. To locate the target at any arbitrary position, the code needs to be able to split a sphere into two volume at MC0, using volume subtraction method.   

In LBNE g4lbne source code, new methods for upstream and downstream physical volume placement, namely LBNEVolumeplacements::PlaceFinalUpstrMultiSphereTarget and LBNEVolumePlacements::PlaceFinalDownstrMultiSphereTarget, were written in parallel with the volume placement methods for simple cylinder and simple box. 

In the original code, the number of graphite fin targets and target position with respect to horn 1 is calculated based on surveyed information entered by user. This will allow users to run the simulation based on actual measurements. LBNEVolumePlacements::SegmentTarget() was branched out do a similar task on SAT. The method is able to adjust target length and position to get an integer number of spheres in total, in upstream and downstream sections. 

The above methods are called after an user interface (UI) card fUseMultiSphereTarget. Another new UI cards in LBNEPlacementMessenger was also defined to take sphere radius as an input. These cards are used in macros files to change the configurations.

The new geometry was verified using plots of Geantino tracks, HepRApp and OpenGL visualization. Once the new geometry has been verified to work, simulations were run on Fermilab computer clusters. Each configuration was run for 500 jobs, each with 100,000 Proton on Target (POT) at beam power 1.2MW. Jobs of the same configuration were merged together and make comparison histograms of neutrino flux at the Far Detector. The neutrino of interest are unoscillated muon neutrino $\nu_{\mu}$, electron neutrino $\nu_e$ and their anti-partners compared to LBNE Nominal design. 

\begin{figure}
\includegraphics[scale=0.3]{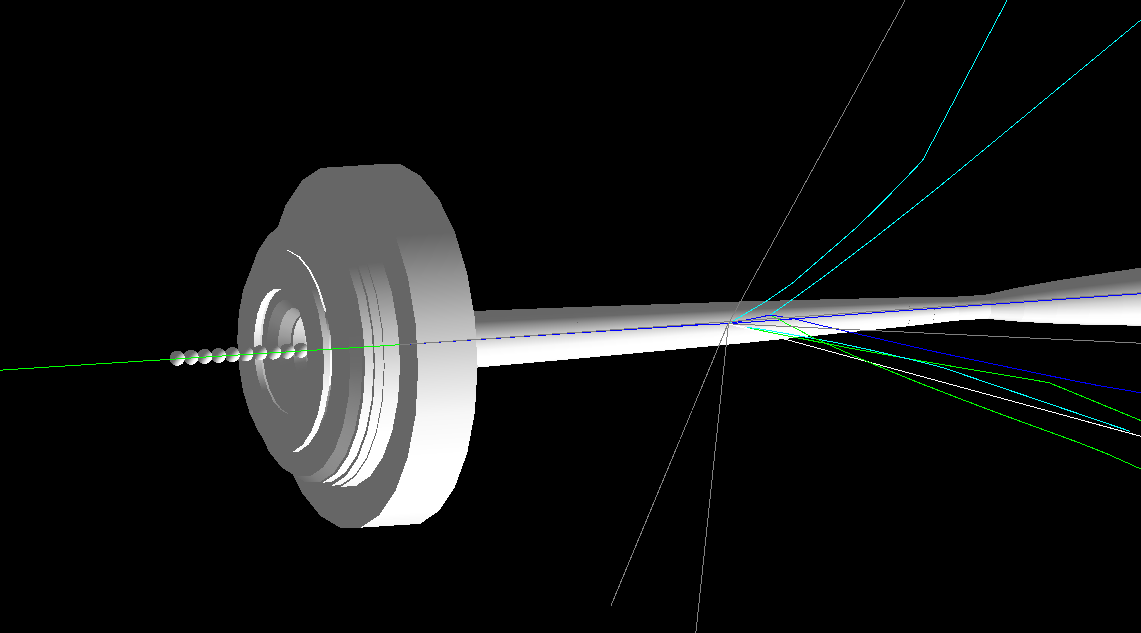} 
\caption{OpenGL visualization of target and inner conductor of horn 1 with a proton on target resulting in an interaction.}
\label{event}
\end{figure}
\section{\label{sec:level1}Result and Analysis}
\begin{figure}
\includegraphics[scale=0.5]{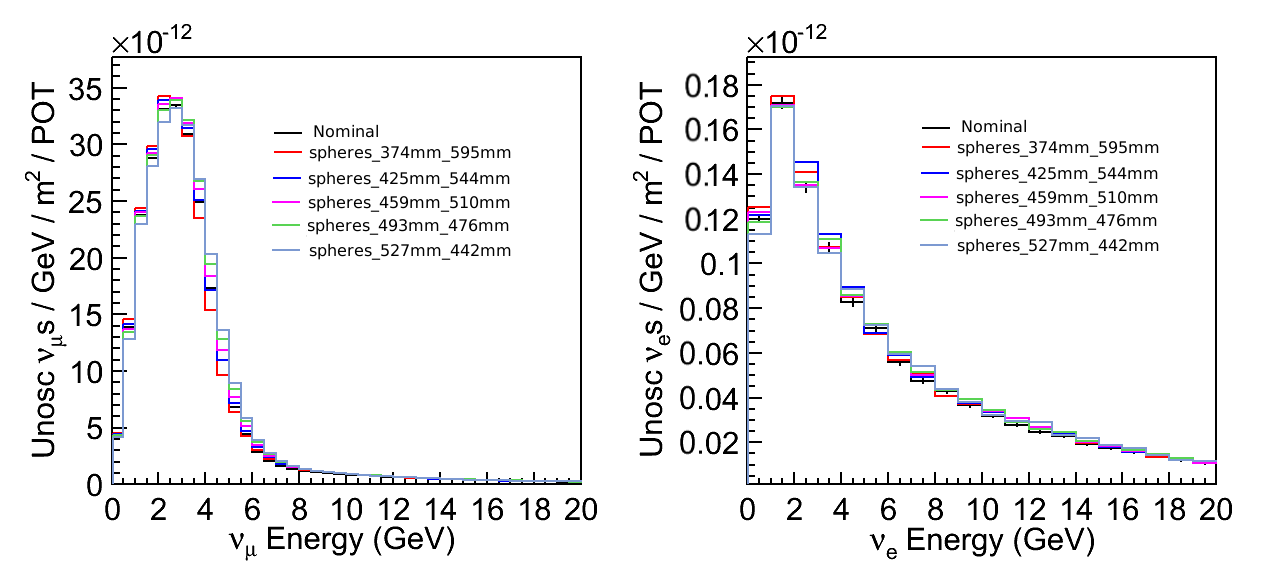} 
\caption{Neutrino flux produced with SAT at the Far Detector. Sphere diameter: 17mm, beam sigma: 17/6mm, target length: 969mm. Legend indicates length of target outside and inside horn 1 in mm, respectively. Nominal plot (black) refers to flux produced by LBNE Graphite target with length inside horn 1 is approximately 504 mm, comparable to the violet plot.}
\label{17_2p8_pos_flux}
\end{figure}

The target was pushed inside horn 1 deeper, using less than 3mm radial clearance (which is the clearance used by LBNE Nominal design). Using target diameter of 17mm and two interaction length of material (969mm) Fig. \ref{17_2p8_pos_flux} shows a higher neutrino flux at 2-3 GeV except for the most retracted target (shown in gray). The higher flux is possibly due to the geometry of the SAT with more than two interaction length at center. 

\begin{figure}
\includegraphics[scale=0.5]{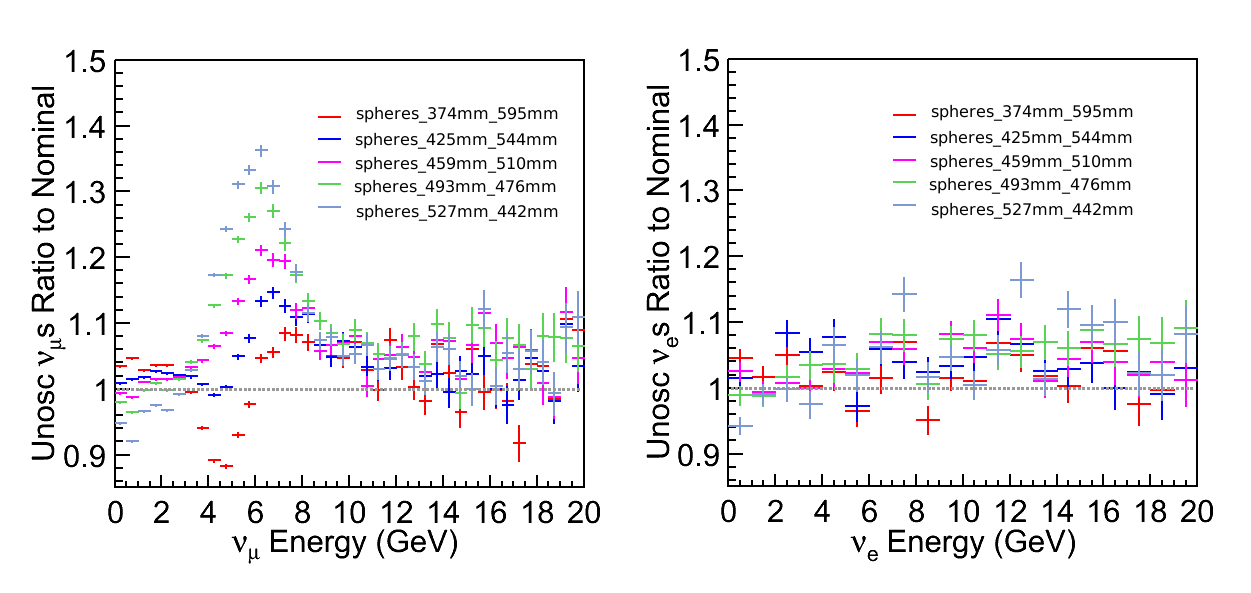} 
\caption{Ratio of neutrino flux produced with SAT over LBNE Nominal target. Sphere diameter: 17mm, beam sigma: 17/6mm, target length: 969mm.}
\label{17_2p8_pos_ratio}
\end{figure}

The ratio plot in Fig. \ref{17_2p8_pos_ratio} shows the gain ratio of flux compared to LBNE Nominal graphite target. The violet plot has a similar target length inside horn 1. To measure the first and second oscillation maximum, the range of interest is 0.5 - 4 GeV. A more favorable flux is shown by the red plot, where the target is pushed deeper inside and flux gain is suppressed at high energy. A sharp transition as the target is retracted from the deepest position (shown in red) by only 51mm (blue plot) begs for more in depth investigation.

\begin{figure}
\includegraphics[scale=0.6]{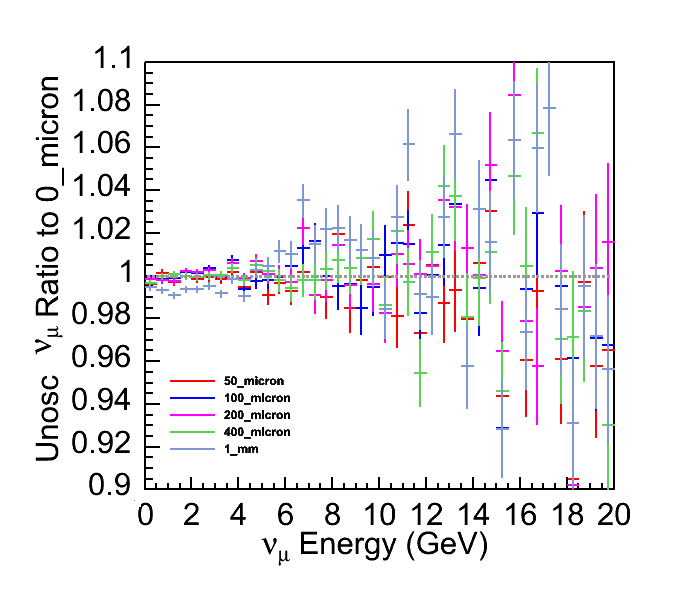} 
\caption{Ratio of neutrino flux produced by SAT at the Far Detector of off-centre beam over on-centre beam. Sphere diameter: 17mm, beam sigma: 17/6mm, target length: 969mm.}
\label{17_2p8_offset_ratio}
\end{figure}

Plot in Fig. \ref{17_2p8_offset_ratio} shows a relatively stable neutrino flux for the chosen beam size of R/3. Flux loss is approximately 1\% when beam offset is 1mm. This result is roughly in agreement with prediction of probability of proton interaction in target in Table. \ref{prob}.

\begin{figure}
\includegraphics[scale=0.6]{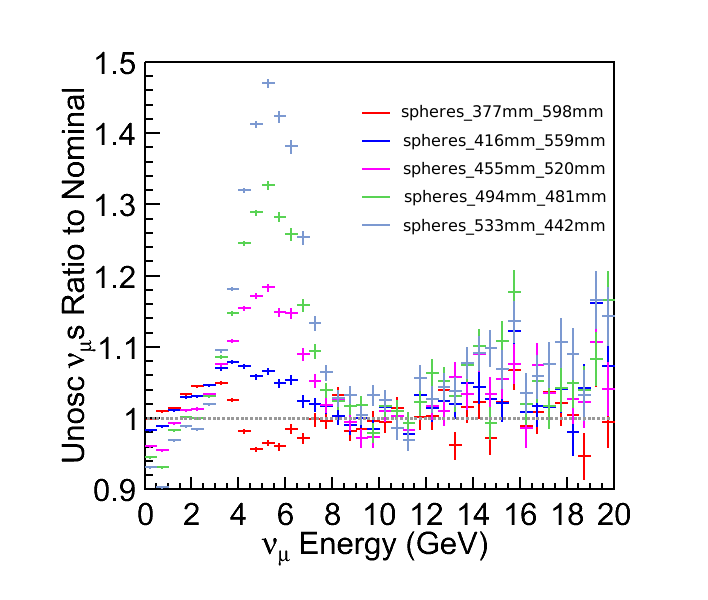} 
\caption{Ratio of neutrino flux produced with 13mm SAT at the Far Detector of off-centre beam over LBNE Nominal target. Sphere diameter: 13mm, beam sigma: 13/6mm, target length: 975mm
}
\label{13}
\end{figure}
\begin{figure}
\includegraphics[scale=0.6]{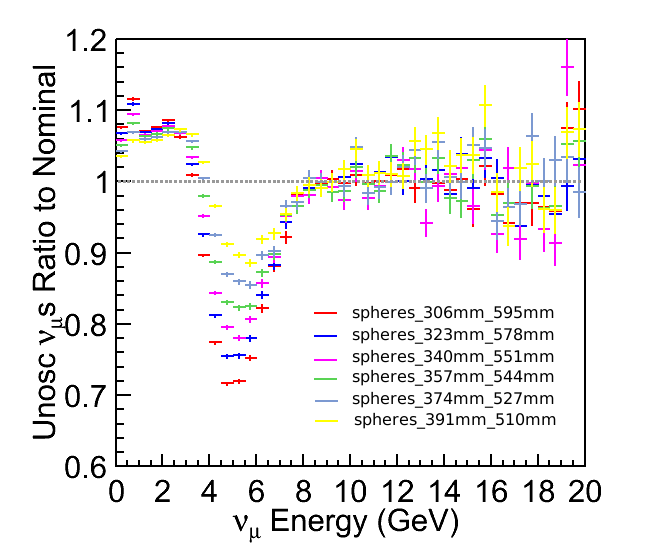} 
\caption{Ratio of neutrino flux produced with SAT with less than two interaction lengths at the Far Detector of off-centre beam over LBNE Nominal target. Sphere diameter: 17mm, beam sigma: 1.7mm, target length: 901mm}
\label{short_target}
\end{figure}

Decreasing sphere diameter to 13mm (which elongates the target to 975mm) shifts the flux gain to higher energy and introduce flux loss at lower energy, shown in Fig. \ref{13}. In most target positions, flux is loss at the first oscillation maximum of 0.8GeV. Flux increases up to 50\% at 6 GeV, which is not of much interest at 1300km.
\begin{figure}
\includegraphics[scale=0.6]{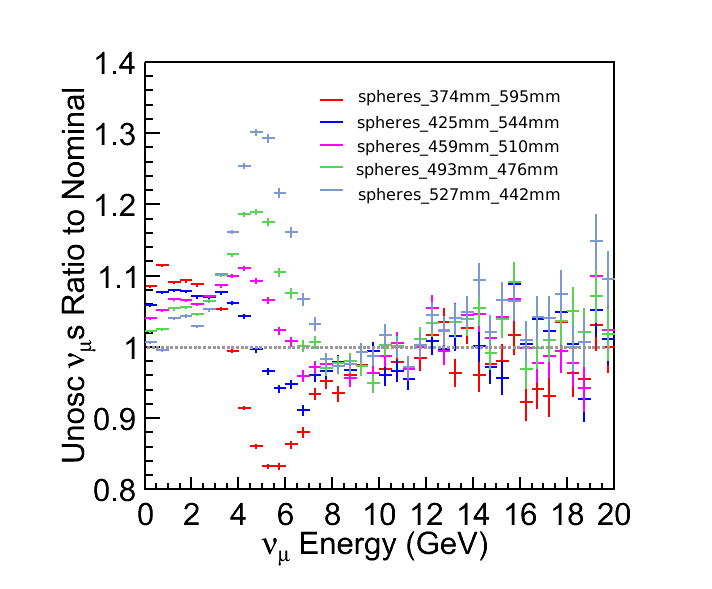} 
\caption{Ratio of neutrino flux produced with SAT at the Far Detector of off-centre beam over LBNE Nominal target. Sphere diameter: 17mm, beam sigma: 1.7mm, target length: 969mm}
\label{17_1p7_pos_flux}
\end{figure}
A smaller beam size of 1.7 mm was also used on 17mm diameter spheres for simulation, result is shown in Fig. \ref{17_1p7_pos_flux}. The flux gain is higher at low energy, compare to the beam size of 2.8mm. A possible explanation is when the beam is more concentrated in the center with longer path length, produced particles traverse through more material and as a consequence, loose energy and produce lower energy neutrino. The plot also shows that, the less target extension outside horn 1, the better flux gain at 0-3.5GeV. This is better demonstrated by the plot in Fig. \ref{short_target}. With the same target length inside horn 1 (shown in red), a shorter target give a better flux scheme: higher gain at 0-3.5GeV, and more flux suppression at after 3.5 GeV. 

\section{conclusion}
The spherical array target with less than two interaction length and a smaller beam sigma of 1.7mm gives higher flux at 0-3.5 GeV and suppresses flux at energy higher than 3.5 GeV. For spherical array target, the beam is relatively stable with less than 2\% flux loss as the beam is offset from target center by 1mm.

A further study of spherical array target should isolate the effect of different parameters and study each separately: material (graphite vs. Beryllium), geometry (fins vs. spherical), total target length and target length inside/outside horn 1.

More engineering study is required to obtain accurate size of Helium can, size and material of helical spacer around target spheres and stress of spherical array target under a beam size of 1.7mm.

The source code also need to be upgraded to move the target to any arbitrarily position and to include the helical spacer. It's also possible to show number of protons interacted in target and interaction location to better understand the effect of target size and geometry.

\section{\label{ack}acknowledgement}
I am grateful to my advisors, Dr. Alberto Marchionni , Dr. Paul Lebrun and Dr. Laura Fields at Fermilab for their support and guidance.


\begin{thebibliography}{9}

\bibitem{lbne}

Milind Diwan, Robert Svoboda, and James Strait. The long-baseline neutrino experiment. {\em European Strategy Preparatory Group}  {\bf 2012}.

\bibitem{volume2}

 LBNE Collaboration. Long-Baseline Neutrino Experiment (LBNE) Project Conceptual Design Report Volume 2: The Beamline at the Near Site. {\em Fermi National Accelerator Laboratory} {\bf 2012}.

\bibitem{ral}
  CJ Densham, O Caretta, TR Davenne,
 MD Fitton, P Loveridge, M Rooney.
  Conceptual Design Study of the LBNE Target and Beam Window, Final Report. {\em STFC Rutherford Appleton Laboratory } {\bf 2010}

\bibitem{t2k}
 T. Nakadaira. Performance and operational feedback of T2K graphite target. {\em T2K Collaboration} {\bf 2012}


\end{thebibliography}
\end{document}